\documentclass[reprint,showpacs,amsmath,amssymb,aps]{revtex4-1}

\usepackage{graphicx}
\usepackage{dcolumn}
\usepackage{bm}

\begin{document}


\title{
Results of a search for daily and annual variations of \\
the $^{214}$Po half-life at the two year observation period
}



\author{E.N.~Alexeyev,$^1$
Yu.M.~Gavrilyuk,$^1$
A.M.~Gangapshev,$^1$
V.V.~Kazalov,$^1$
V.V.~Kuzminov,$^1$
S.I.~Panasenko,$^2$
S.S.~Ratkevich$^2$
}
\affiliation{$^1$ Baksan Neutrino Observatory INR RAS, Russia}
\affiliation{$^2$ V.N.Karazin Kharkiv National University, Ukraine}

\date{\today}%

\begin{abstract}
The brief description of installation TAU-2 intended for long-term monitoring of the half-life value
$\tau$ ($\tau_{1/2}$) of the $^{214}$Po is presented.
The methods of measurement and processing of collected data are reported.
The results of analysis of time series values $\tau$ with different time step are presented.
Total of measurement time was equal to 590 days.  Averaged value of the $^{214}$Po half-life was
obtained $\tau=163.46\pm0.04$ $\mu$s. The annual variation with an amplitude $A=(8.9\pm2.3)\cdot10^{-4}$,
solar-daily variation with an amplitude $A_{So}=(7.5\pm1.2)\cdot10^{-4}$, lunar-daily variation with
an amplitude $A_L=(6.9\pm2.0)\cdot10^{-4}$ and sidereal-daily variation with an amplitude
$A_S=(7.2\pm1.2)\cdot10^{-4}$ were found in a series of $\tau$ values.
The maximal values of amplitude are observed at the moments when the projections of the installation
Earth location velocity vectors toward the source of possible variation achieve its maximal magnitudes.
\end{abstract}

\maketitle

\section{\label{sec:intro}Introduction}
At the last time in works intended to search for limits of the realization of the decay constant conservation law, a level of sensitivity not less than $2\cdot10^{-4}$ was reached for several radioactive isotopes. In the work \cite{a1} the authors showed an amplitude of a possible annual variation of  the $^{198}$Au half-life ($T_{1/2}=2.69445$ days), that was measured with the relative uncertainty of $\pm7\cdot10^{-5}$ does not exceeds  $\pm2\cdot10^{-4}$ of the central value. Variations with periods from several hours up to one year were excluded at the level of $9.6\cdot10^{-5}$ (95\% C.L.)
during the measurements of the  $^{137}$Cs half-life ($T_{1/2}=10942$ days) in the \cite{a2}. The annual variation was excluded at the level of $8.5\cdot10^{-5}$ (95\% C.L.).
Variations of an activity with periods of 3-150 days were excluded at the level of $2.6\cdot10^{-5}$ (99.7\% C.L.)
during the measurement of the  $^{40}$K activity in the \cite{a3}. It was shown that an amplitude of the annual variation does not exceeds of $6.1\cdot10^{-5}$ (95\% C.L.).
Variations of an activity with periods less then one year were excluded at the level of $4\cdot10^{-5}$ during the measurement of the  $^{232}$Th activity in the \cite{a3}.

A count rate of the detector recording the source radiation was a subject of investigations in the all mentioned works. A high sensitivity of the measurements was reached by using of a relatively high count rate ($\sim10^3$~s$^{-1}$), of a control and a stabilization of conditions of the measurements and by using of additional arrangements for shield of the set-ups from outer background.

At the same time, the evidences of a presence of the annual variations of different effects caused by a radiation of the investigated isotope are cited in a series of articles. For example, characteristics of the annual  variations of the count rates of the detectors used to many years measurements of the $^{32}$Si and $^{226}$Ra sources activities discussed in the work \cite{a4}. The amplitudes of variations are equal to $\sim1\cdot10^{-3}$. The authors have examined possibilities of an appearing of such variations as a result of seasonal variations of the detector's characteristics or the one of an annual modulation of the isotope's decay rates themselves under the action of an unknown factor depending of the Earth-Sun distance. Results of continuous measurements of the decay rates of the $^{108}$Ag, $^{133}$Ba, $^{137}$Cs, $^{152}$Eu, $^{154}$Eu, $^{85}$Kr, $^{226}$Ra and $^{90}$Sr sources made in the Physikalisch-Technische Bundesanstalt (PTB) was discussed in the works \cite{a5, a6}. Statistically significant annular variations with the amplitudes of $(6.8-8.8)\cdot10^{-4}$ were observed in the data for all this isotopes. The authors made a comparison between spectral power functions obtained from the data and the observed radial oscillations of the San's surface. However, the researchers from the PTB point in the \cite{a7} at the laboratory factors capable to cause similar variations and warn against hasty conclusions about possible new physical phenomena.

It is clear that any conclusions about a possible new physical effect could be made after the complete exclusion of variations caused by influence of the known terrestrial geophysical, climatic and meteorological factors on the source-detector couple count rate. Not all such factors could be detected and took into account during the measurement and data processing. For example, an annual variation with the amplitude of $(4.5\pm0.8)\cdot10^{-5}$ was found as a result of a processing of the data collected at 500 days of Earth's surface measurement with $^{40}$K source in the \cite{a3}. It was found that this variation corresponds completely to the known annual variation of the cosmic rays intensity and could be explained by a cosmic rays background events contribution to the total detector's count rate. A variation with the $\sim300$ days period and $4\cdot10^{-5}$ amplitude was found in the data collected at 480 days in the underground measurement with the $^{232}$Th source.  It was found that this variation correlate with a variation of a daily averaged dead time per event and could be explained by a modulation of the RC circuit providing the shaping time of the amplifier.

The weak point of the experiments aimed to monitor a stability of a controlled radiation count rate is the high sensitivity to the similar variations of the measurement conditions. It seems that this shortcoming became unimportant for the decay constant determination based on direct registration a life time of nucleus between a birth and a decay.  This  methodic was realized by us in the \cite{a8} for the $^{214}$Po which decays with 164.3 $\mu$s half-life \cite{a9} by emitting the 7.687 MeV $\alpha$-particle. This isotope appears mainly in the exited state ($\sim87$\%)
in the $^{214}$Bi $\beta$-decay. Half-lives of the exited levels does not exceed 0.2 ps \cite{a10} and they discharge instantly with regard to the scale of  the $^{214}$Po half-life. Energies of the most intensive $\gamma$-lines are equal to 609.3 keV (46.1\%
per decay), 1120 keV (15.0\%) and 1765 keV (15.9\%).
So, the  $\beta$-particle and $\gamma$-quantum are emitted at the moment of a birth of $^{214}$Po nuclear (start) and the $\alpha$-particle are emitted at the decay moment (stop). Measurement of "start-stop" time intervals allows one to construct decay curve at an observation time and to determine the half-life from it's shape. The $^{226}$Ra source ($T_{1/2}=1600$ years) was used as a generator of $^{214}$Bi nuclei which arise in the decay sequence of the mother isotope.

The direct measurement of a nuclear life time allows one moreover to study the radioactive decay law itself. The theoretical models discussed in the \cite{a11, a12} predict that the decay curves could deviate from the exponential law in the short- and very long-time regions of the time scale. The theoretically predicted \cite{a13, a14, a15} so called quantum Zeno effect consisting in a slowing down of the decay rate in a case of constant observations at the decaying object presents a special interest. Experimentally Zeno effect was proved \cite{a16} in repeatedly measured two-level system undergoing Rabi transitions, but not observed in spontaneous decays.

A limitation of the annual variation amplitude was set at the level of $3.3\cdot10^{-3}$ at the first stage of our measurements \cite{a8}. Factors limiting a sensitivity were revealed and ways of it's optimization were designed. The improvements of the set-ups, methods of measurements and data processing, results of an analysis of possible sources of systematic errors were described in the work \cite{a17}. The annular variation with an amplitude of $(6.9\pm3.0)\cdot10^{-4}$ and the solar-daily variation with an amplitude of $(10.0\pm2.6)\cdot10^{-4}$ were obtained as a result of a processing of the data collected at 480 days of the measurements.

A brief description of the used installation and results of a broadened analysis of the data collected at 590 days are given in the present work.

\section{Method of measurements}

The TAU-2 set-up consists of the two scintillation detectors D1 and D2 used in the work. The D1 was made of two glued discs of a plastic scintillator (PS) with the 18 mm diameter (d) and 0.8 mm thickness (h). A thin transparent circular bag glued of two 2.5 $\mu$m lavsan layers placed between the discs. A radium spot is deposited preliminary in the center of inner surface of a one lavsan circle. The D1 registers $\beta$-particles from the $^{214}$Bi decays and $\alpha$-particles from the $^{214}$Po decays. The massive detector D2 made of NaI(Tl) crystals  intended for the $\gamma$-quanta detection. Two NaI(Tl) crystals (d=150 mm, h=150 mm) placed by ends one to another with the 10 mm gap. The D1 is placed into a gap between D2a and D2b. The light collection is fulfilled from a lateral side of the PS disc installed into deep narrow well with a reflecting wall. The measurements are carried out in the low background room of the underground laboratory DULB-4900 of the BNO INR RAS at the depth of 4900 m of water equivalent \cite{a18} in the additional shield made from Pb (15 cm).

	A registration of the pulses in the set-up is carried out by the two-channel digital oscilloscope La-n20-12PCI which is inserted into a personal computer (PC). Pulses are digitized with 6.25 MHz frequency (160 ns/channel).  The DO pulse recording starts by a signal from the D2 which registered $^{214}$Bi decay's $\gamma$-quanta. A D2 signal opens a record of a sequence with 655.36 $\mu$s total duration where first 81.92 $\mu$s time is a "prehistory" and the last 573.44 $\mu$s is a "history". Duration of a "history" exceeds the three $^{214}$Po half-lives.

	A scintillation detector D1 in the TAU-2 has a relative $\alpha/\beta$ light output $\sim0.1$ \cite{a19}.
As result the pulses from the $\alpha$- and $\beta$-particles have the comparable amplitudes.
This circumstance was used to a preliminary selection of the "useful" events by the "on-line" program prepared the data for a PC recording. It is obtained a number of pulses in the D1 and D2 channels and its delays. The pulses have tested at correspondence to the "correct event" criterions such as 1) the only one pulse in the D2 channel, 2) presence of the prompt coinciding pulse in the D1 channel, 3) presence of the only one delayed pulse in the "history" of the D1 channel and 4) absence of any pulses in the "prehistory" of the D1. The frames do not correspond to the "correct event" criterions are rejected.  A "correct" event appearing time, pulse amplitudes and its appearing time are recorded into the PC memory. This information allows one to "off line" process the data for the different regions of pulse amplitudes. A count rate of the "right" events is equal to $\sim12$~s$^{-1}$. A rate of an information accumulation is equal to $\sim25$ Mb$\cdot$day$^{-1}$.

\section{Search for long duration variations of the TAU-2 data}

\begin{figure}[pt]
\includegraphics*[width=2.5 in,angle=270.]{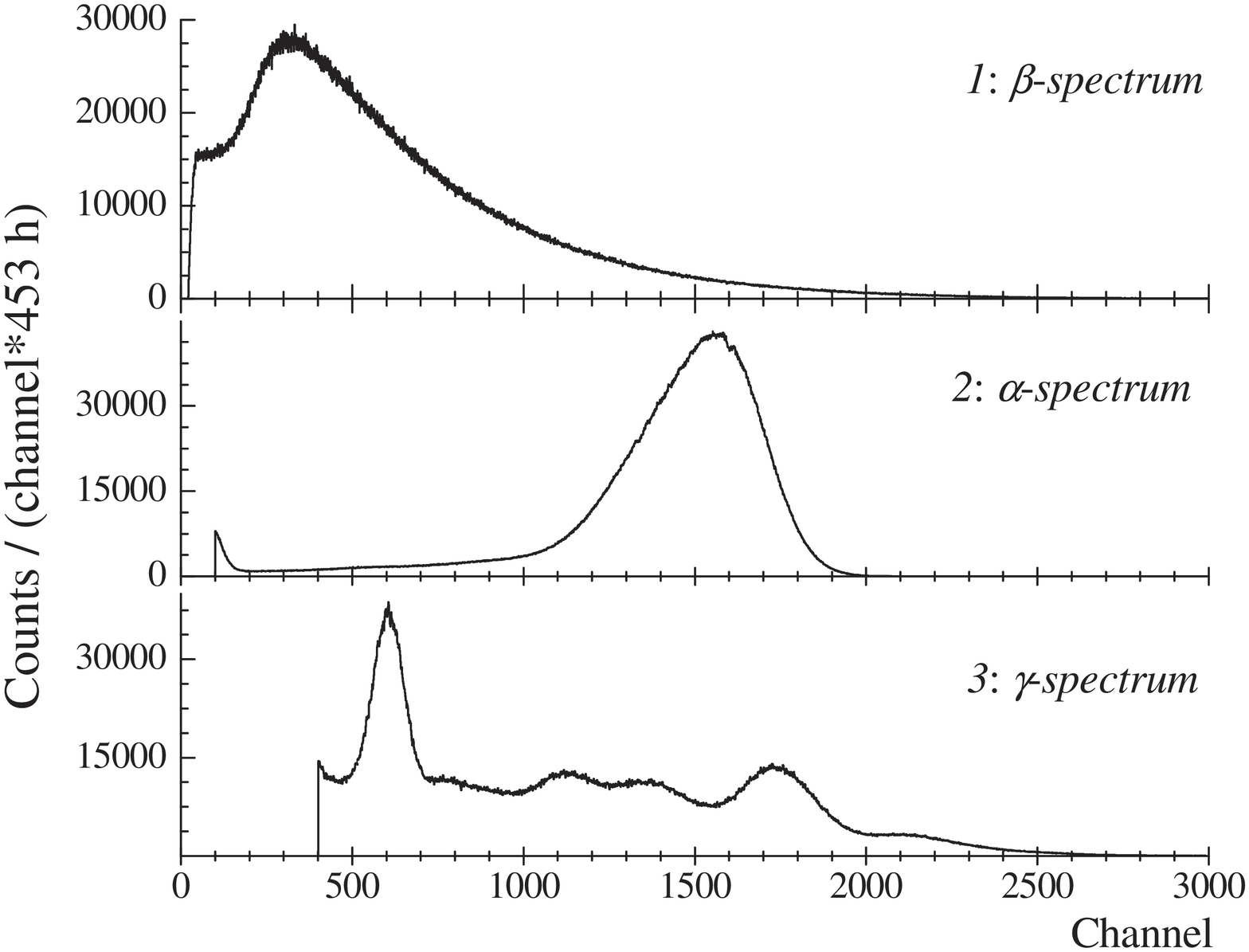}
\caption{\label{fig1} Spectra of the coincided in 573.44 $\mu$s delay of pulses from the detector D1
($\beta$-spectrum \emph{1}) and D2 ($\gamma$-spectrum \emph{3}) and the $\alpha$-spectrum \emph{2} of one of the first pulse D1 to install TAU-2.
}
\end{figure}

\begin{figure}[pt]
\includegraphics*[width=2.2 in,angle=270.]{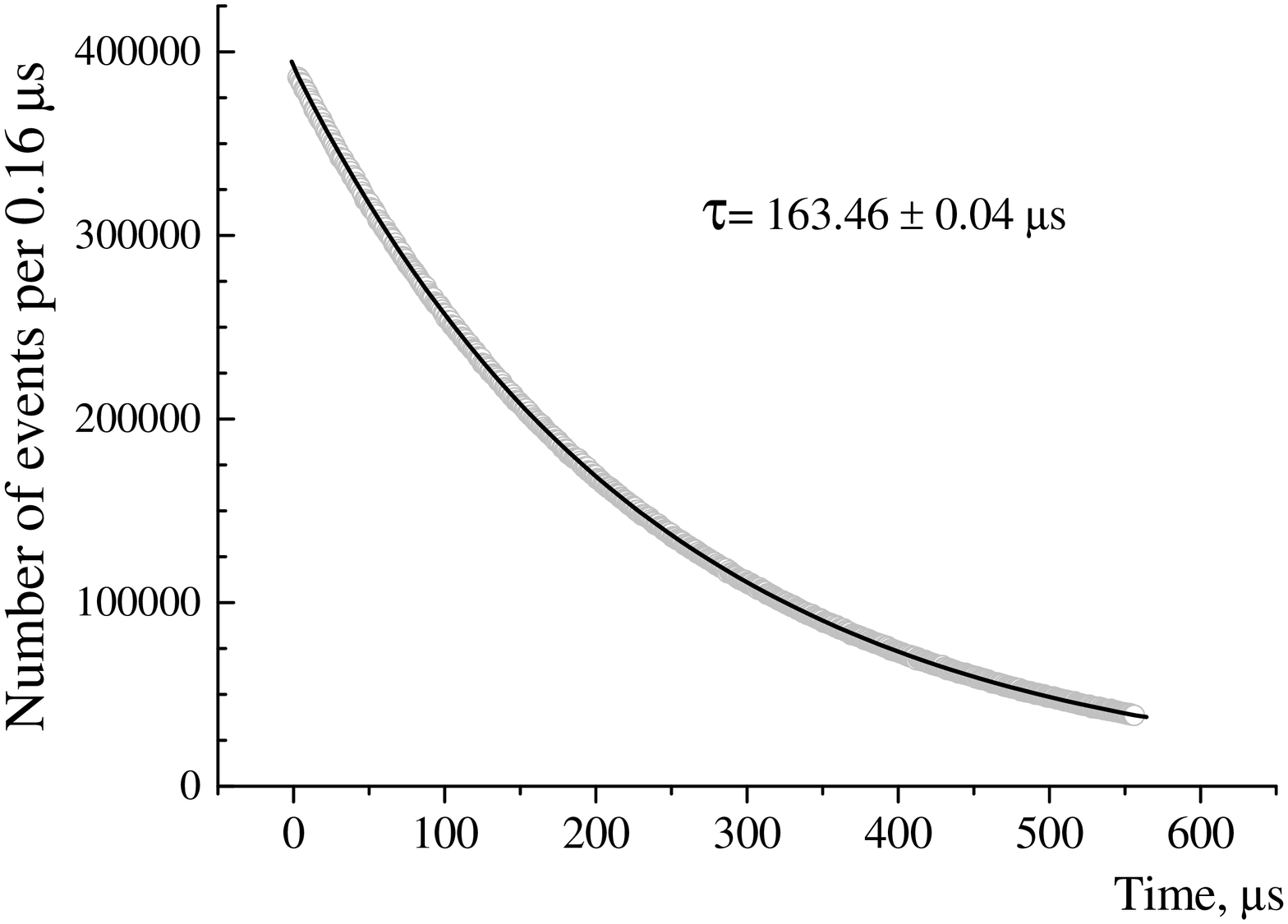}%
\caption{\label{fig2}
The start-stop delays distribution obtained in 590 days at the TAU-2.
The decay-curve of the $^{214}$Po with $\tau=163.45\pm0.04$ $\mu$s.}
\end{figure}

The spectra of the $\beta$-pulses (spectrum 1) and $\alpha$-pulses (spectrum 2) of the D1 detector and of the $\gamma$-pulses (spectrum 3) of the D2 detector corresponding to the selection conditions mentioned above are shown in Fig.\ref{fig1}.
The peak at the channel $\sim1600$ on the spectrum 2 is formed by the 7.69 MeV $\alpha$-particles. A total time of the data collection is equal to 730 days at the period October 2012 – October 2014. A decay curve constructed for the total data set is shown in Fig.\ref{fig2}.
 A value of the $\tau$ obtained from this data is equal to $163.45\pm0.04$ $\mu$s. The $\tau$-value equal to ($164.58\pm0.29$ (stat.)$\pm0.10$ (syst.)) $\mu$s  was measured at the Gran Sasso in a recent work \cite{a20}. These two values are compatible in the $1\sigma$ limits.

\begin{figure}[pt]
\includegraphics*[width=2.35 in,angle=270.]{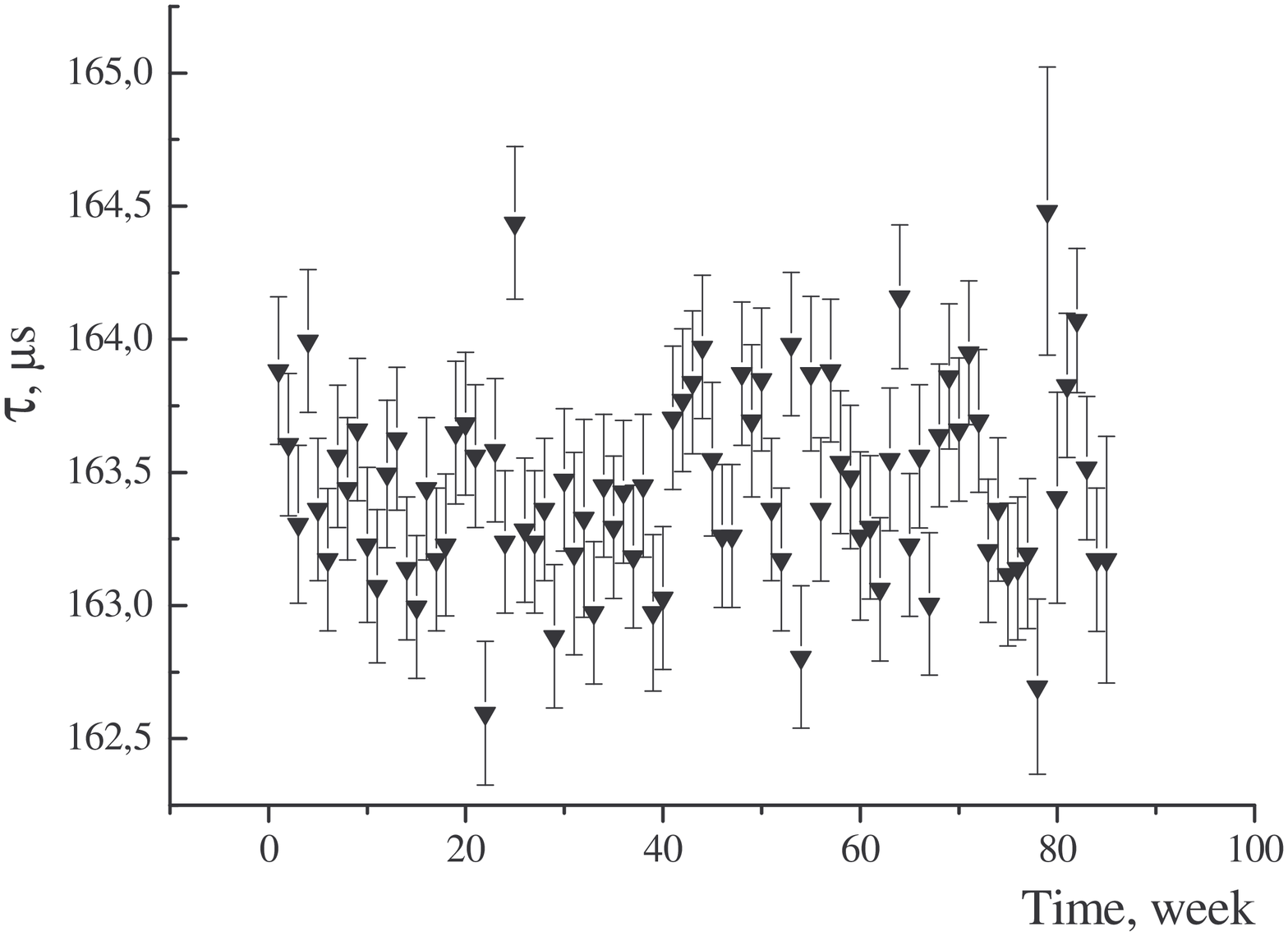}%
\caption{\label{fig3}
Dependence in time of the $\tau$-value obtained at the TAU-2 with the week step.}
\end{figure}

The continuous in time data set was divided at portions of the equal length to search for possible time variations. The decay curve has constructed for the each partition data set and the continuous in time sequence of the $\tau$-values with the specified time step has found.  The time dependence of the $\tau$-values with the week time step is shown in Fig.\ref{fig3}.
The $\tau$-values were defined for the 3.2-560 $\mu$s delay-time region by means of a $\chi^2$-approximation of the decay-curves collected at one week each with an exponential function
\begin{equation} \label{eq1}
y=a \cdot exp(-ln(2)\cdot t/\tau)+b.
\end{equation}

\begin{figure}[pt]
\includegraphics*[width=2.35 in,angle=270.]{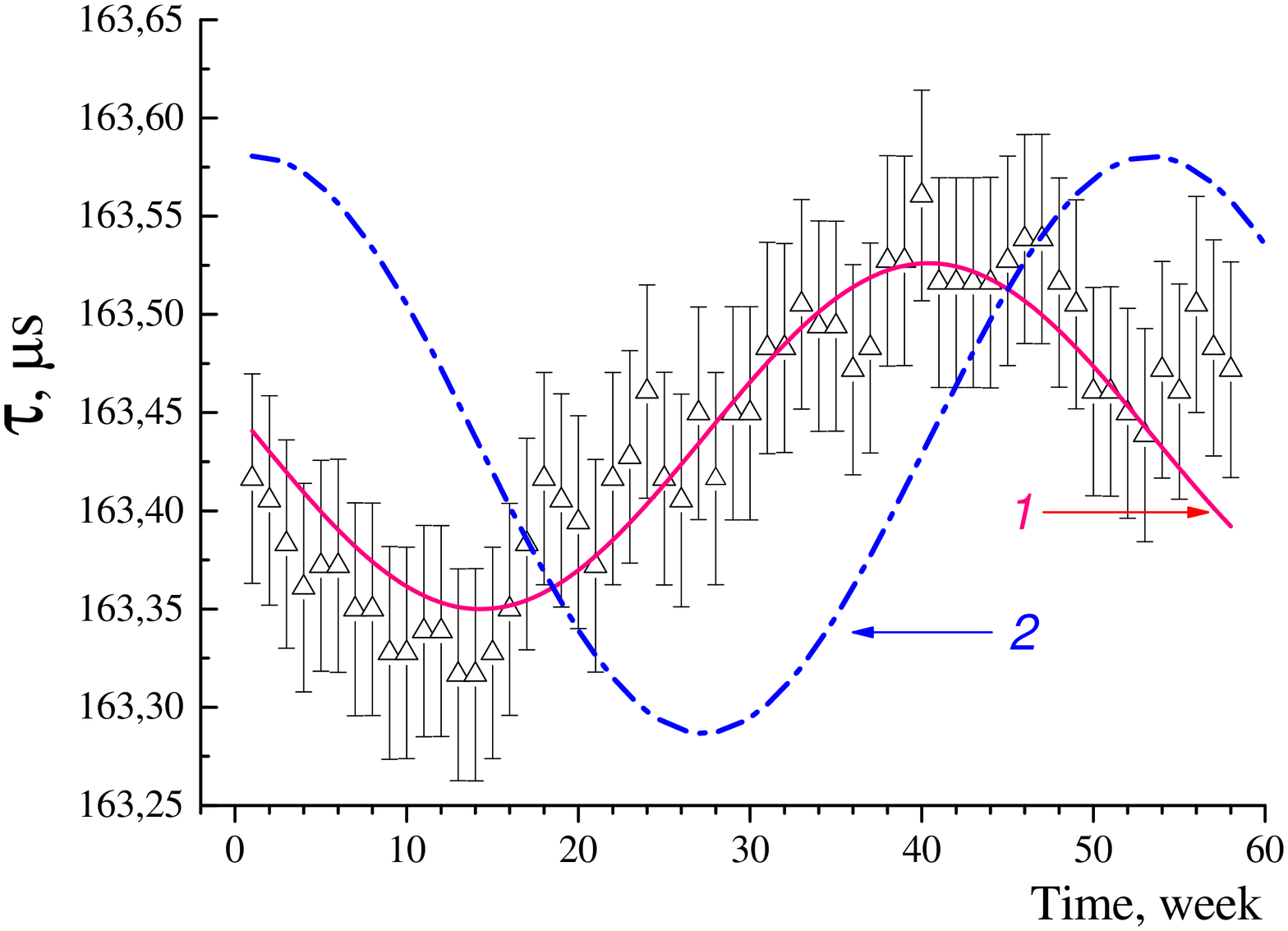}%
\caption{\label{fig4}
Distribution in time of the $^{214}$Po half-life  values obtained by the "moving-average method".
Curve \emph{1} – approximation  by function (\ref{eq2}) with the parameters: $A=5.4 \cdot 10^{-4}$, $\omega =2\pi / 365$ d$^{-1}$, $\varphi=83$ d (since the 1st of January).
Curve \emph{2} – approximation  by function (\ref{eq2}) with the parameters: $A=8.9 \cdot 10^{-4}$, $\omega =2\pi / 365$ d$^{-1}$, $\varphi=174$ d.
}
\end{figure}

The moving-average method (moving summation) was used to search for a possible time variations of the $\tau$-values. A time interval with the duration equal to about 0.5 of the expected period is choose to search for any harmonic component and the $\tau$-value is determine for this interval. Than, the interval have shifted for a one step and the procedure is repeated. The 0.5 year interval and the one week step were chosen to search for the annular variations. A decay curve was constructed for the data collected at 0.5 year and the $\tau$-value was determined for it. Than, the interval have shifted for the 1 week and a determination was repeated. The result of the analysis is presented on the Fig.\ref{fig4}.

\begin{figure}[pt]
\includegraphics*[width=2.2 in,angle=270.]{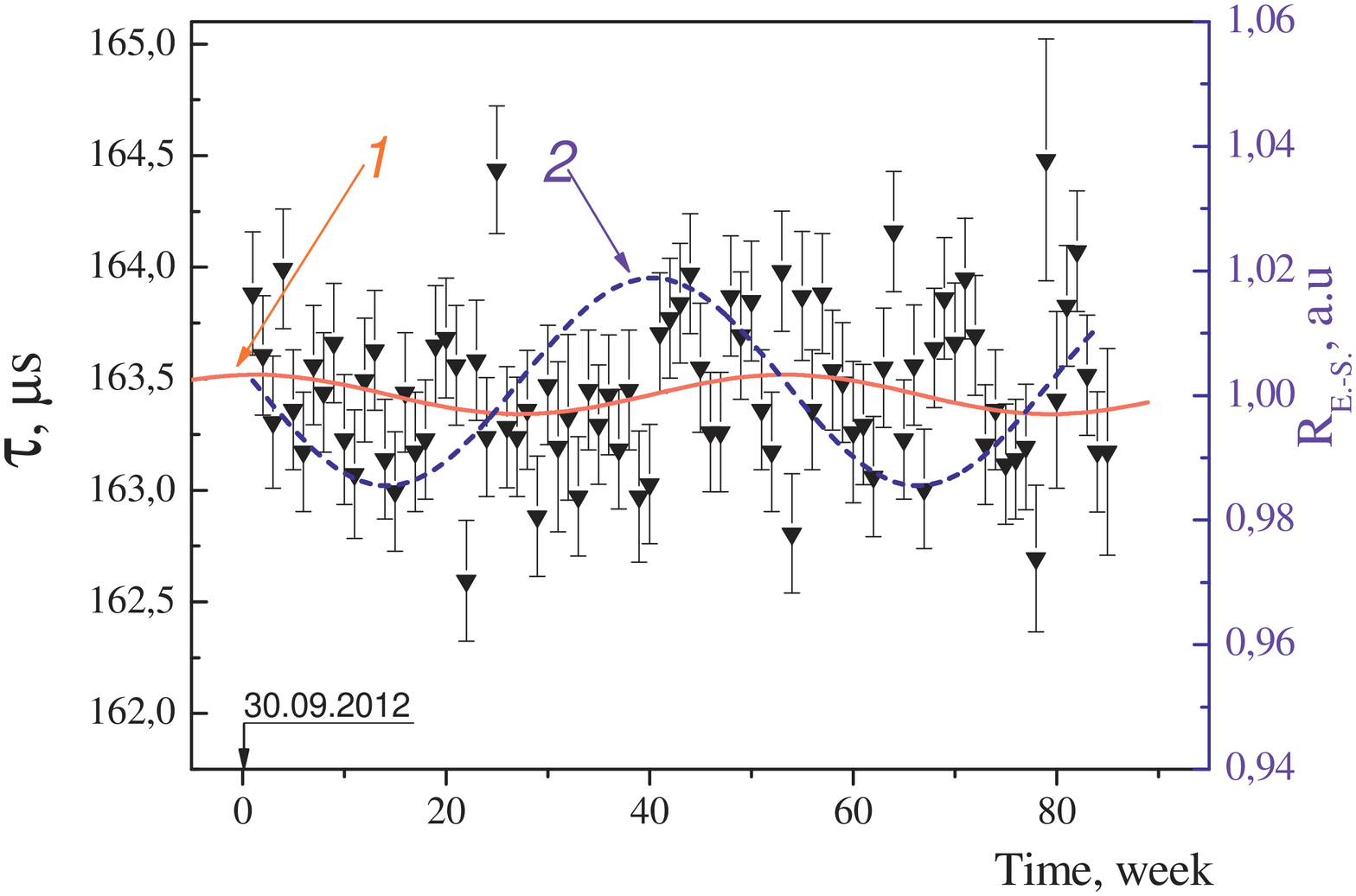}%
\caption{\label{fig5}
Distribution in time of the $^{214}$Po half-life values for  one week data sets.
Curve \emph{1} – approximation  by function (\ref{eq2}) with the parameters: $A=8.9 \cdot 10^{-4}$, $\omega =2\pi / 365$ d$^{-1}$, $\varphi=174$ d.
Curve \emph{2} - time dependence of  E.-S. distance (right scale).
}
\end{figure}

Harmonic component have presented in the data as it seen from the figure. The  approximation dependence
\begin{equation} \label{eq2}
\tau(t)=\tau_0 \cdot [1+A \cdot sin(\omega \cdot(t+\varphi))]
\end{equation}
where $A = 5.4 \cdot 10^{-4}$, $\omega = 2\pi / 365$ d$^{-1}$, $\varphi = 83$ d (since the 1st of January) obtained from the data by means of $\chi^2$-method is shown by red color on the Fig.\ref{fig4} too.
It is easy to show that the original dependence of the week data set has the same period (1 year), the amplitude large at $\pi / 2$ time and shifted at 0.25 year (0.5 of the sliding interval) than the approximation one. The annual wave
\begin{equation*}
\tau (t) = \tau_0 \cdot [1+8.9 \cdot 10^{-4} \cdot sin(\omega \cdot (t+174))]
\end{equation*}
obtained by such a way is shown on Fig.\ref{fig5} together with the week data set of the $\tau$-values.

It is follows from the consideration that the periodic component with the 1 year period and the amplitude of
\begin{equation*}
A=(8.9 \pm 2.3) \cdot 10^{-4}
\end{equation*}
is presented in the data. The maximum of the periodic function is observed at the $\sim$October 22. Phases of the obtained period and the period of the Earth-Sun distance variation differ at 3 months. Therefore, obtained periodic variation could not be explained by the distance changing. However, a parameter of the Earth velocity relative to the Sun is exists in connection with the Earth-Sun (E.-S.) distance annular variation. A dependence of this parameter is shown of Fig.\ref{fig5} (curve 2).
\begin{figure}[pt]
\includegraphics*[width=2.15 in,angle=270.]{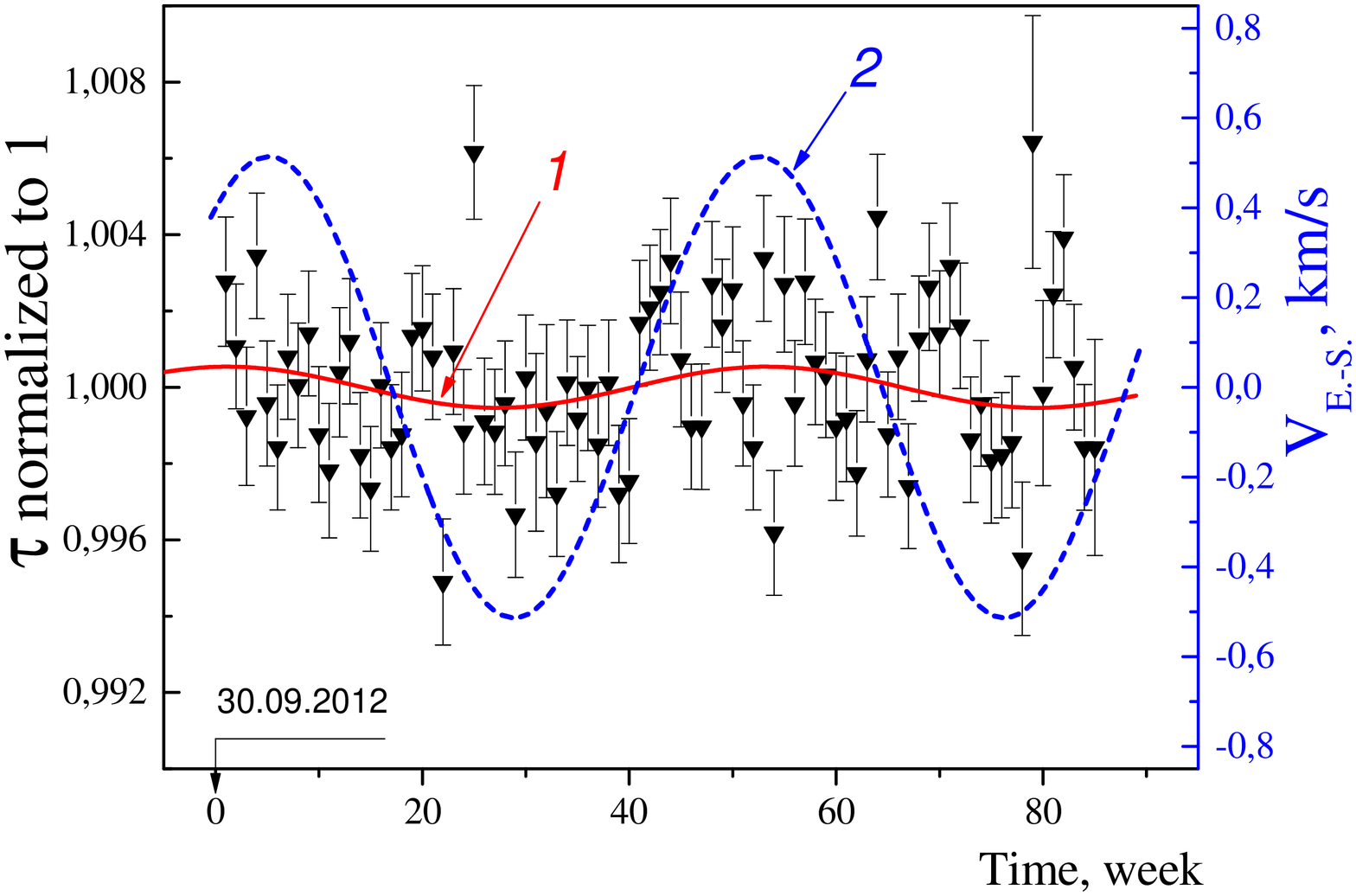}%
\caption{\label{fig6}
Normalized to 1 of distribution in time of the $^{214}$Po half-life  values for a week data sets.
Curve \emph{1} – approximation  by function (\ref{eq2}) with the parameters: $A=8.9 \cdot 10^{-4}$, $\omega =2\pi / 365$ d$^{-1}$, $\varphi=174$ d.
Curve \emph{2} - time dependence of  E. to S. velocity.
}
\end{figure}
The obtained annular wave of $^{214}$Po half-life (Fig.\ref{fig6}, curve 1) is coincides in phases with the E.-S. velocity dependence within the $\pm 1$ week accuracy.

\begin{figure}[pt]
\includegraphics*[width=2.35 in,angle=270.]{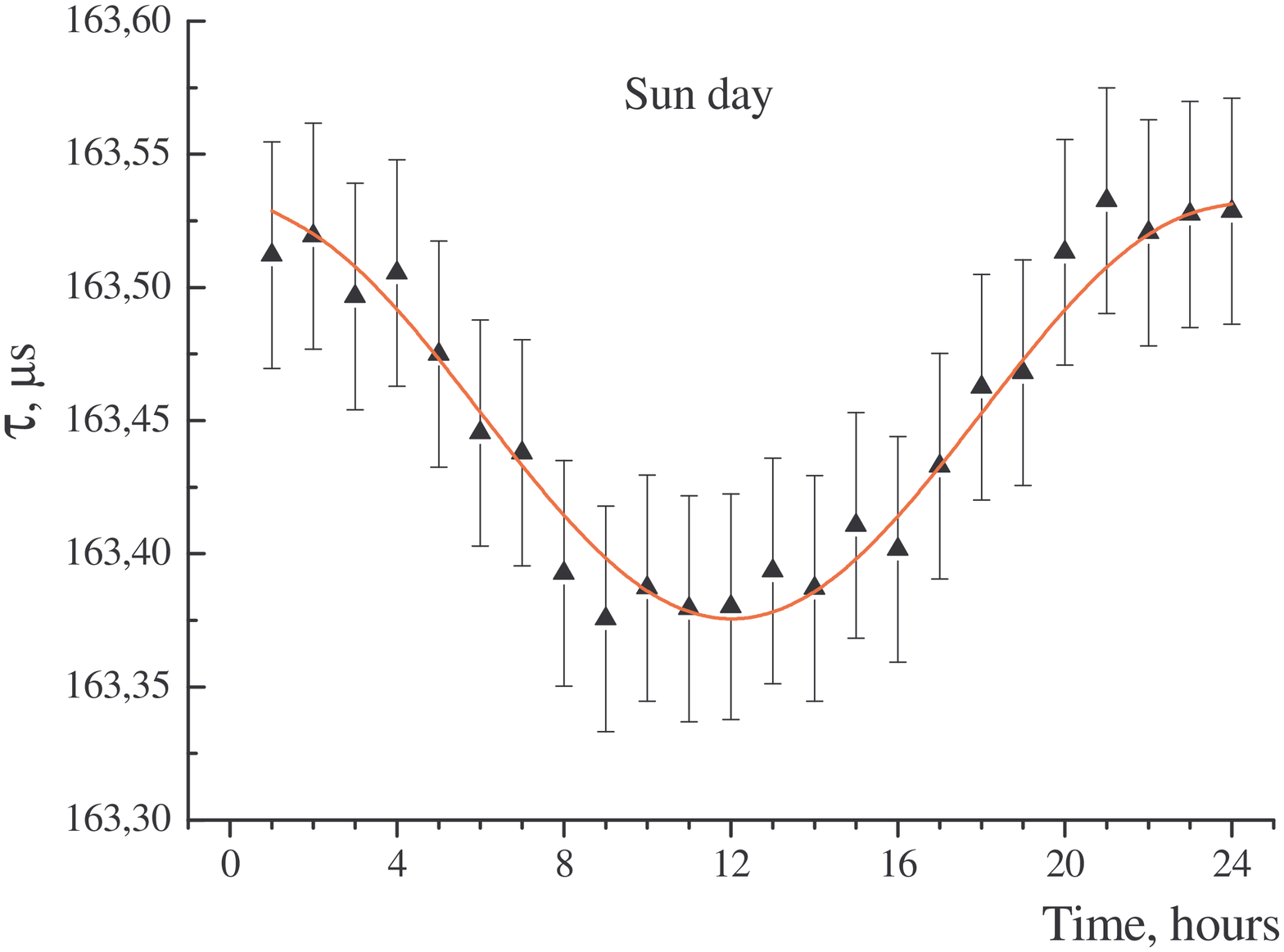}%
\caption{\label{fig7}
Solar-daily variation of half-life of $^{214}$Po obtained by means of moving-average method.
The solid curve - approximation of function (\ref{eq2}) with the parameters: $A=4.8 \cdot 10^{-4}$, $\omega =2\pi / 24$ h$^{-1}$, $\varphi=-6$ h.
}
\end{figure}

\begin{figure}[pt]
\includegraphics*[width=2.35 in,angle=270.]{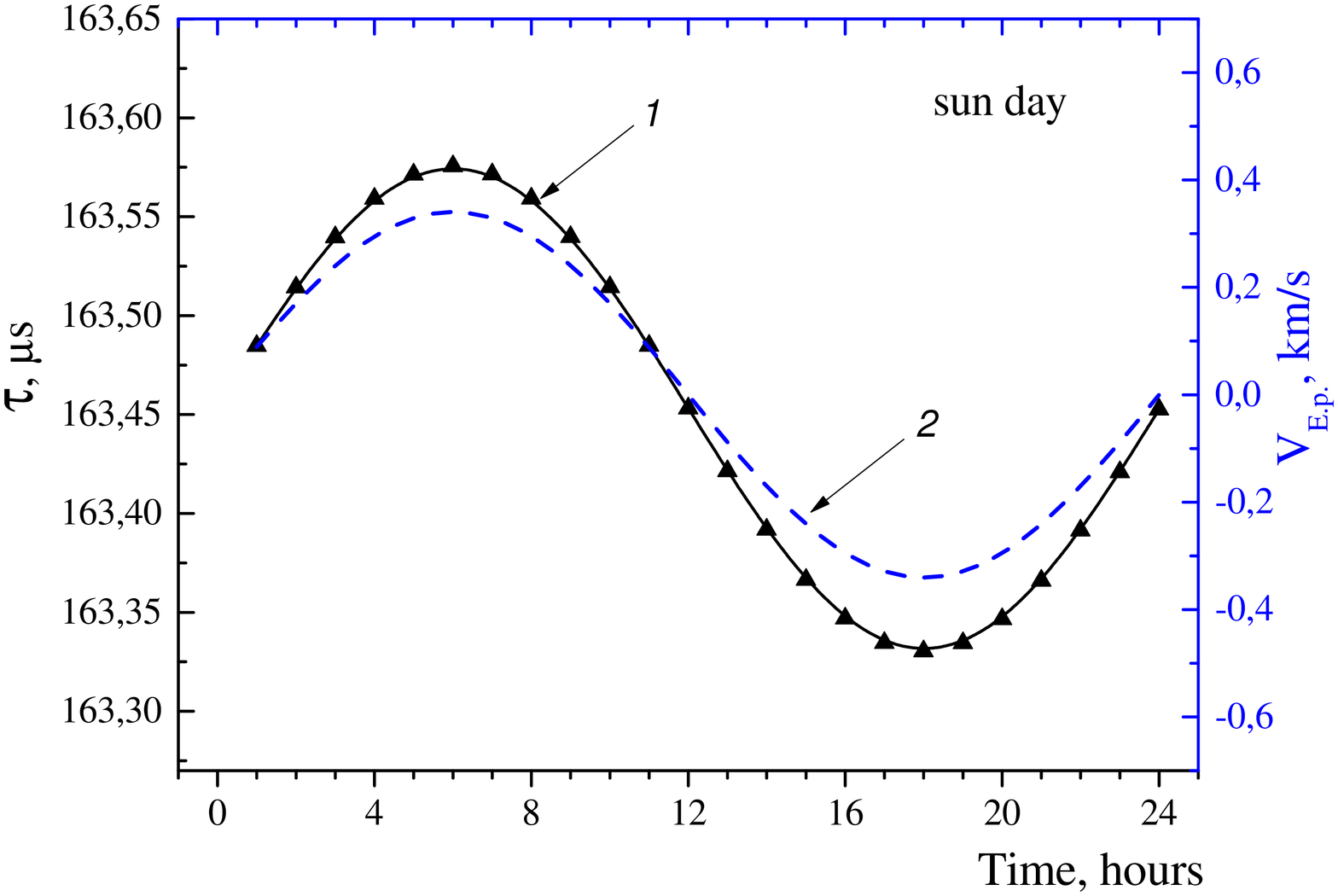}%
\caption{\label{fig8}
Curve \emph{1} - Solar-daily variation of half-life of $^{214}$Po approximated by
function (\ref{eq2}) with the parameters: $A=7.5 \cdot 10^{-4}$, $\omega =2\pi / 24$ h$^{-1}$, $\varphi=0$ d.
Curve \emph{2} - Earth surface point velocity relative to the Sun due to the Earth rotation.
}
\end{figure}

The described above method was used to search for the daily variation in the solar, lunar and sidereal times. A day's duration was divided on 24 hours in the each case. The durations of the lunar and sidereal days in the solar time are equal to 24 h 50 min 28.2 s and 23 h 56 min 4.09 s correspondingly \cite{a21}. The 12 h duration period was chosen as an averaging interval. All events registered in the 0-12 hour interval at the total period of the measurements were used to construct the decay curve. A half-life value determined from this curve. The interval was moved for the 1 hour and procedure was repeated after that. The result of a search for the daily variation in the solar time is shown on the Fig.\ref{fig7}
with the parameters ($A=4.8 \cdot 10^{-4}$, $\omega =2\pi / 24$ h$^{-1}$, $\varphi=-6$ h) of the approximation of function (\ref{eq2}). The daily variation is described good enough by this sine function as it seen from the Fig.\ref{fig7}.
A reconstructed sought for dependence
\begin{equation*}
\tau (t) = \tau_0 \cdot [1+7.5 \cdot 10^{-4} \cdot sin((2 \pi / 24) \cdot t)]
\end{equation*}
is shown on the Fig.\ref{fig8} (curve 1) together with a dependence of the Earth surface point velocity relative to the Sun due to the Earth rotation (curve 2).
Their phases are coincide in the range of $\pm0.5$ hour. The amplitude of a solar daily variation is equal to
$A_{\rm So}=(7.5\pm1.2)\cdot 10^{-4}$.

\begin{figure}[pt]
\includegraphics*[width=2.35 in,angle=270.]{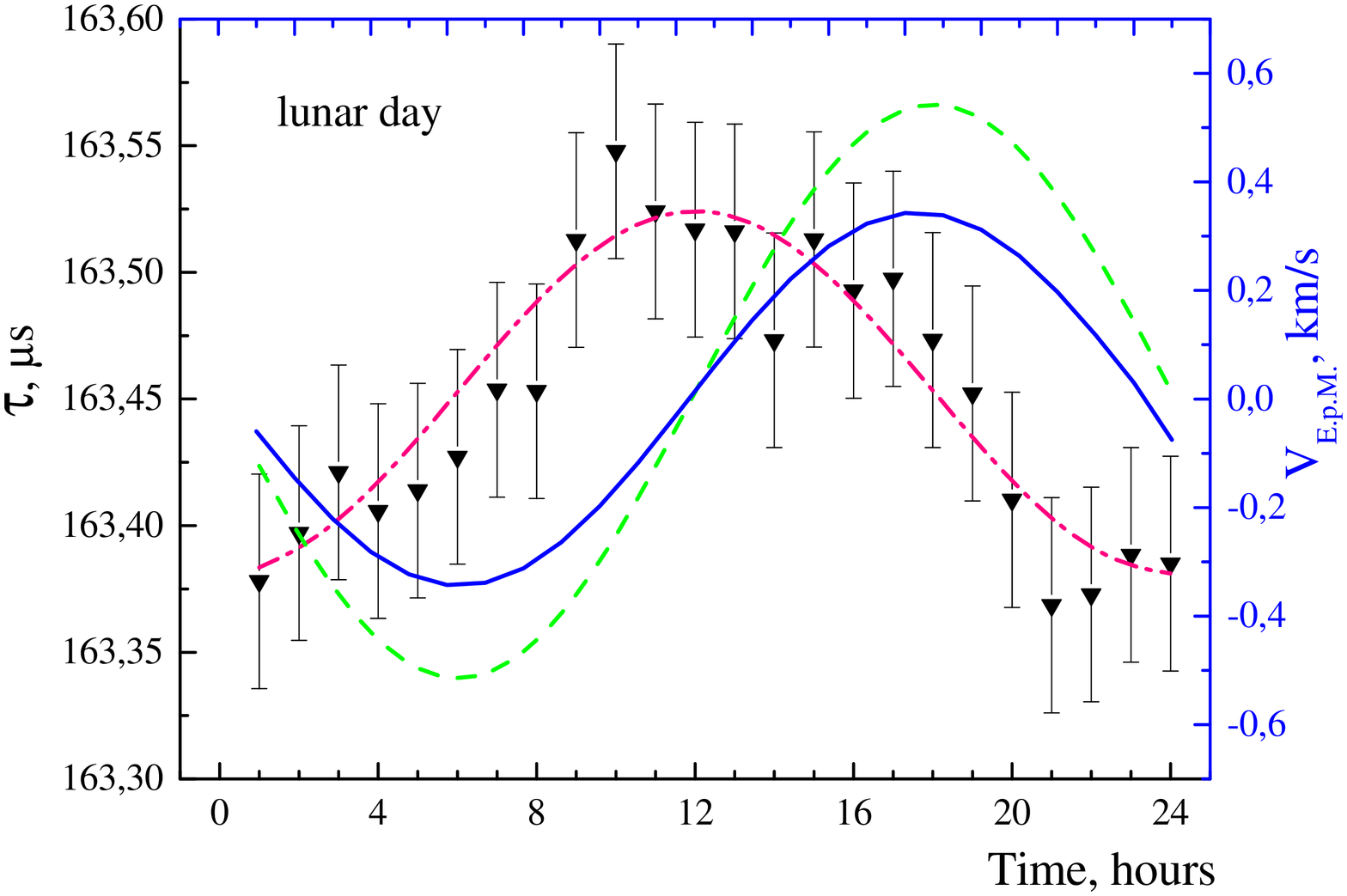}%
\caption{\label{fig9}
Lunar-daily variation of half-life of $^{214}$Po obtained by means of moving-average method.
The data set starting point is 24:00, 30 September 2012.
Dash dot curve - approximation function eq.(\ref{eq2})
with the parameters ($A=4.4 \cdot 10^{-4}$, $\omega =2\pi / 24$ h$^{-1}_{ld}$, $\varphi=6$ h$_{ld}$).
Dash curve - sought Lunar-daily variation $\tau(t) = \tau_0 \cdot [1+6.9\cdot 10^{-4} \cdot sin((2\pi/24) \cdot (t+12))]$.
Solid curve - Earth surface point velocity (V$_{\rm E.p.M/}$) relative to the Moon due to the Earth rotation.
}
\end{figure}

The result of a search for the daily variation $\tau(t)$  in the lunar time is shown on the Fig.\ref{fig9} (curve 1) with the parameters of the approximation function (\ref{eq2}):
\begin{equation*}
  A = 4.4\cdot10^{-4}, ~\omega = 2\pi/24 ~{\rm  h}_{ld}^{-1}, ~\varphi=6 ~{\rm h}_{ld},
\end{equation*}
where h$_{ld}$ – lunar day hour. The reconstructed sought for dependence
\[\tau(t) = \tau0 \cdot[1+6.9\cdot10^{-4}\cdot sin((2\pi/24) \cdot(t+12))]\]
is shown on the Fig.\ref{fig9} (curve \emph{2}) together with a dependence of the Earth surface point velocity relative to the Moon due to the Earth rotation (curve \emph{3}). Phases of the curves (\emph{2}) and (\emph{3}) are coincide in the range of $\pm0.5$ hour. The amplitude of lunar daily variation is equal to
\[A_{\rm L}=(6.9\pm2.0)\cdot10^{-4}.\]

\begin{figure}[pt]
\includegraphics*[width=2.35 in,angle=270.]{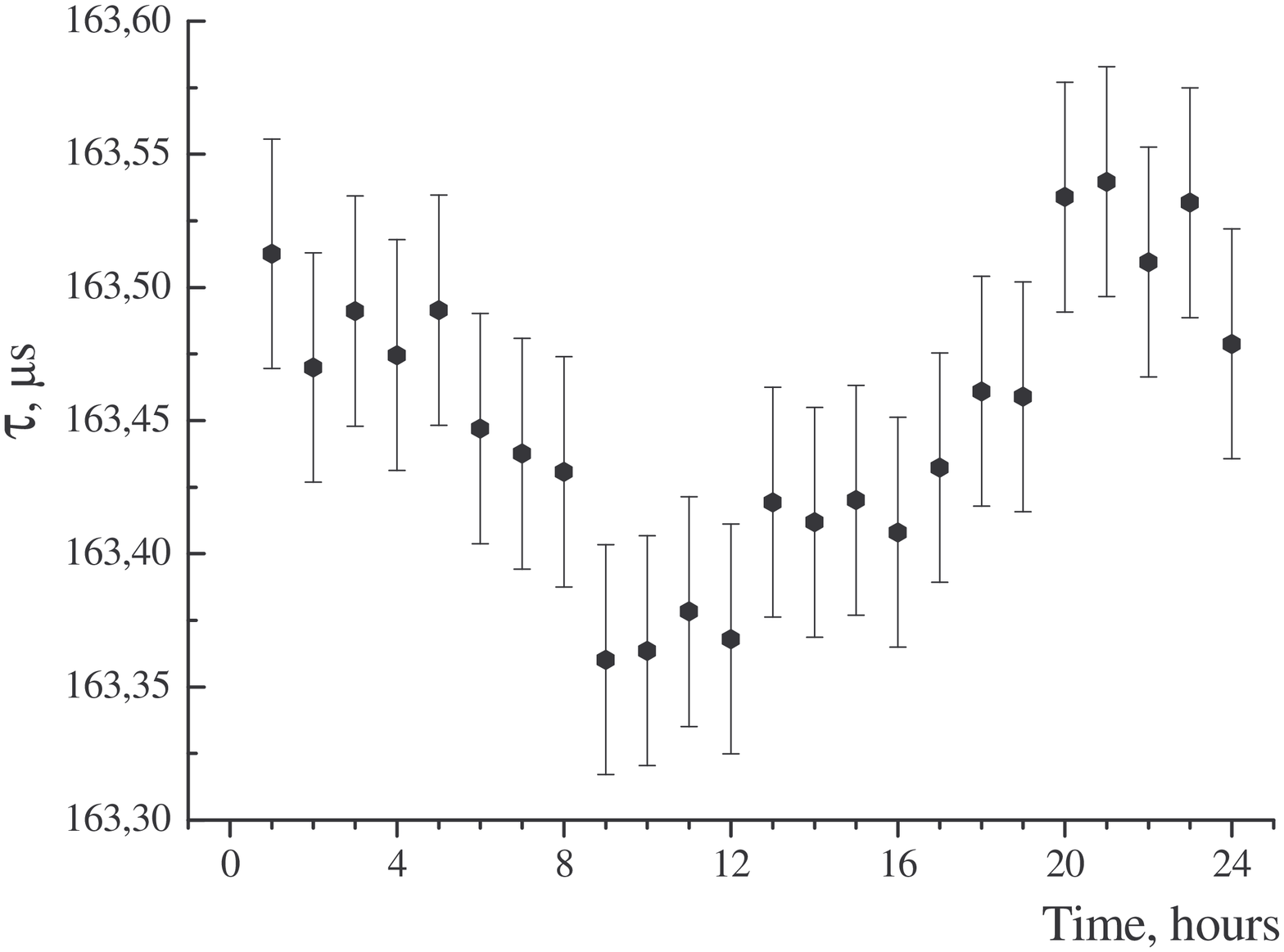}%
\caption{\label{fig10}
Lunar-daily variation of half-life of $^{214}$Po obtained by means of moving-average method.
Starting point shifted for 15 days (15 October 2012).
}
\end{figure}

It was done a test of a real presence of the lunar daily variation in the data set. A starting point of the data summation was shifted at 15 days.
A phase of the variation should shifts at $\sim12$ hours because of a difference in a time duration of a solar and lunar days [(50~min~28.2~s)$\cdot 15 \approx 12$ h]. The result is shown on Fig.\ref{fig10}.
A phase had shifted at 12 hours as it was waited for a real lunar daily variation.

\begin{figure}[pt]
\includegraphics*[width=2.35 in,angle=270.]{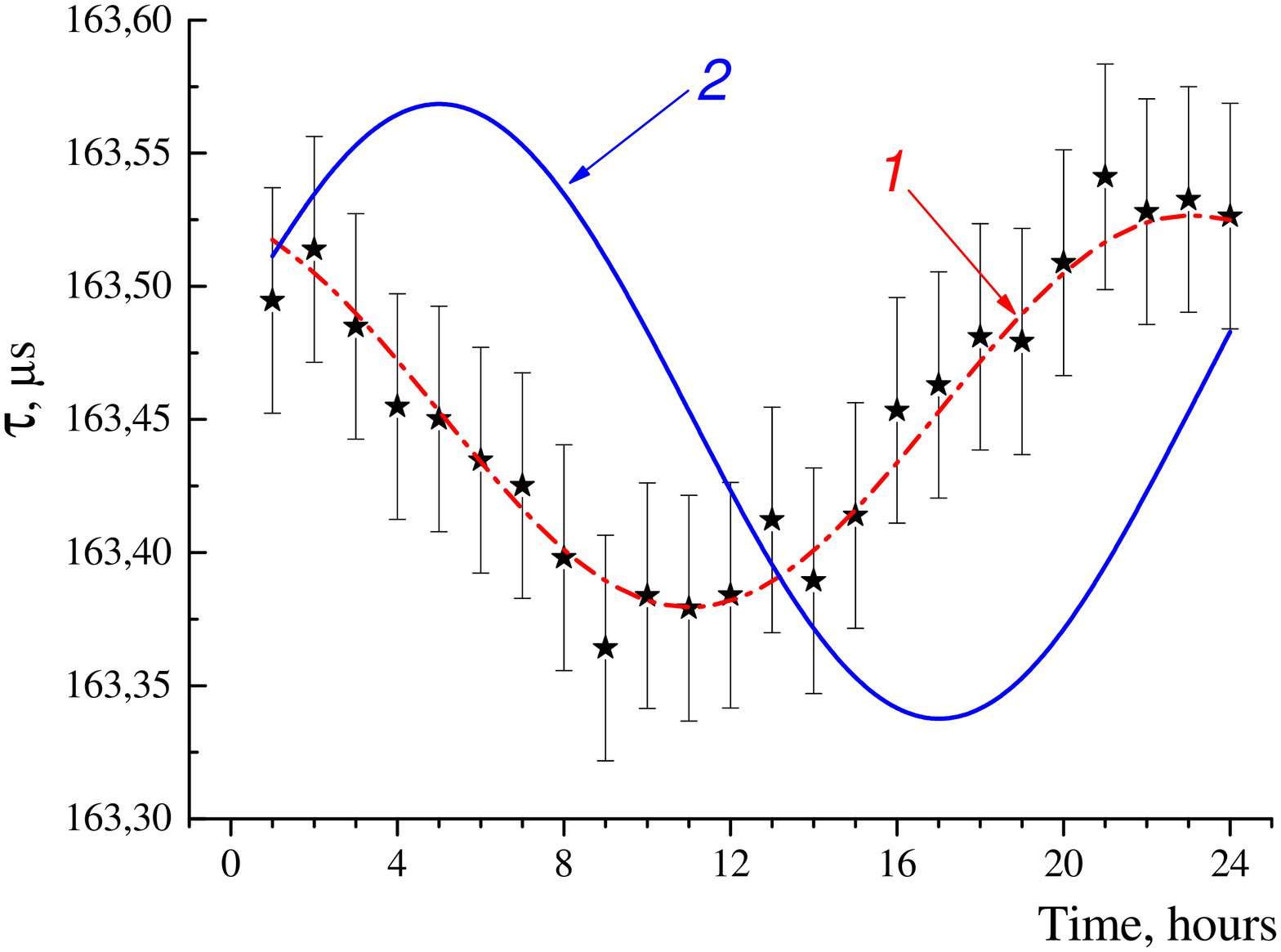}%
\caption{\label{fig11}
Sidereal-daily variation of half-life of $^{214}$Po obtained by means  of moving-average method.
Curve \emph{1} - function eq.(\ref{eq2}) with the parameters ($A=4.6 \cdot 10^{-4}$, $\omega =2\pi / 24$ h$^{-1}_{sd}$, $\varphi=-6$ h$_{sd}$).
Curve \emph{2} - sought SD variation
$\tau(t) =\tau_0\cdot [1+7.2 \cdot 10^{-4} \cdot sin((2\pi/24) \cdot t)].$
}
\end{figure}

The result of a search for the daily variation $\tau(t)$  in the sidereal time is shown on the Fig.\ref{fig11} (curve 1) with the parameters of the approximation function (\ref{eq2}):
\begin{equation*}
  A = 4.6\cdot10^{-4}, ~\omega = 2\pi/24 ~{\rm  h}_{ld}^{-1}, ~\varphi=-6 ~{\rm h}_{ld},
\end{equation*}
where h$_{sd}$ – sidereal day hour. The reconstructed sought for dependence

\begin{equation}  \label{eq3}
\tau(t) = \tau0 \cdot[1+7.2\cdot10^{-4}\cdot sin((2\pi/24) \cdot t)]
\end{equation}
is shown on the Fig.\ref{fig11} (curve 2) too. The amplitude of a sidereal daily variation is equal to \[A_{\rm S}=(7.2\pm1.2)\cdot 10^{-4}.\]

\begin{figure}[pt]
\includegraphics*[width=2.35 in,angle=270.]{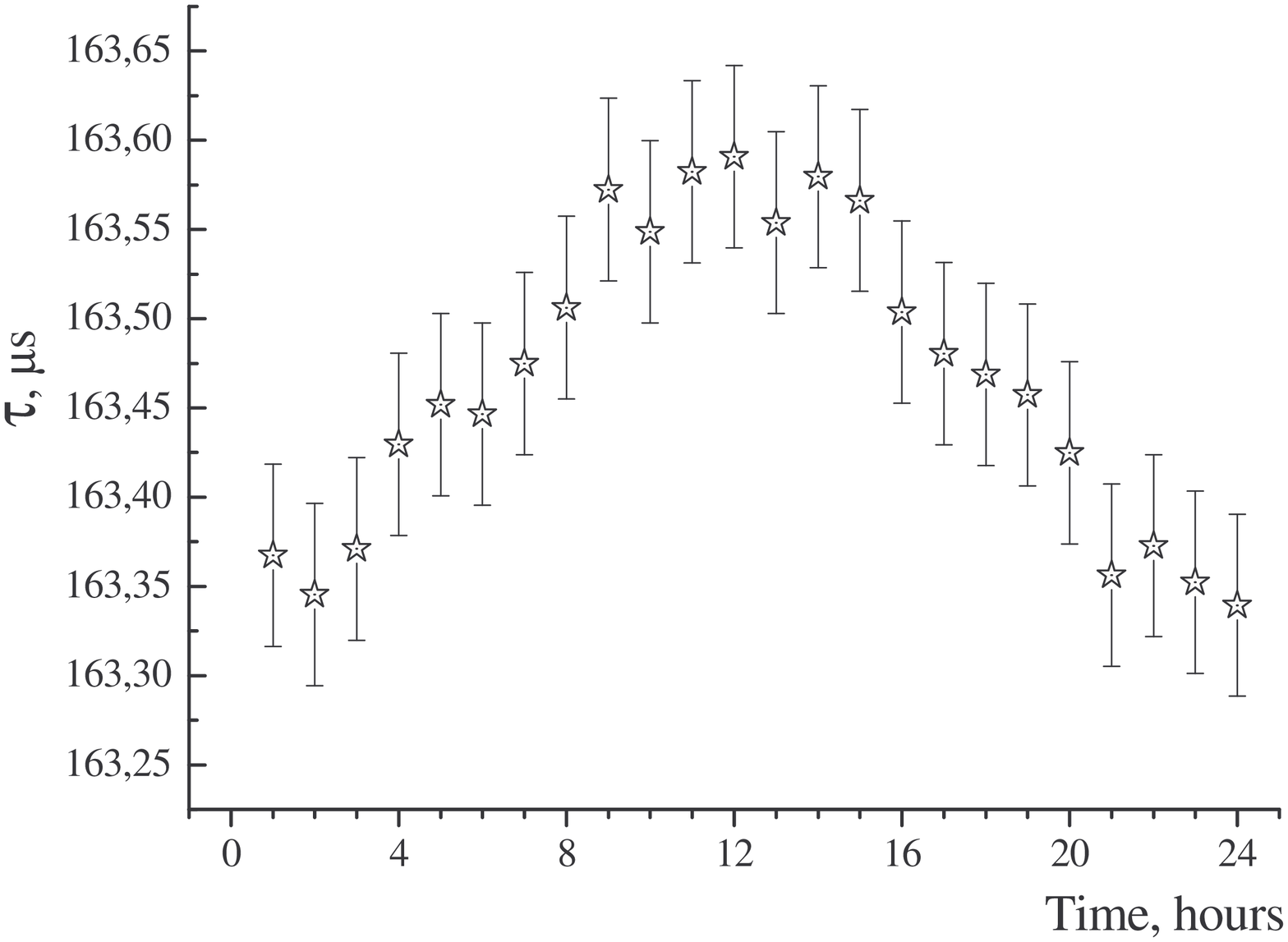}%
\caption{\label{fig12}
Sidereal-daily variation of half-life of $^{214}$Po obtained by means of moving-average method.
Starting point shifted for 182 days.
}
\end{figure}

It was done as test of a real presence of the sidereal daily variation in the data set. A starting point of the data summation was shifted at 182 days. A phase of the variation should shifts at $\sim12$ hours because of difference in a time duration of solar and sidereal days
[(3~min~55.9~s)$\cdot 182 \approx 12$ h].
The result is shown on Fig.\ref{fig12}. A phase had shifted at 12 hours as it was waited for a real sidereal daily variation.

\begin{figure}[pt]
\includegraphics*[width=2.35 in,angle=270.]{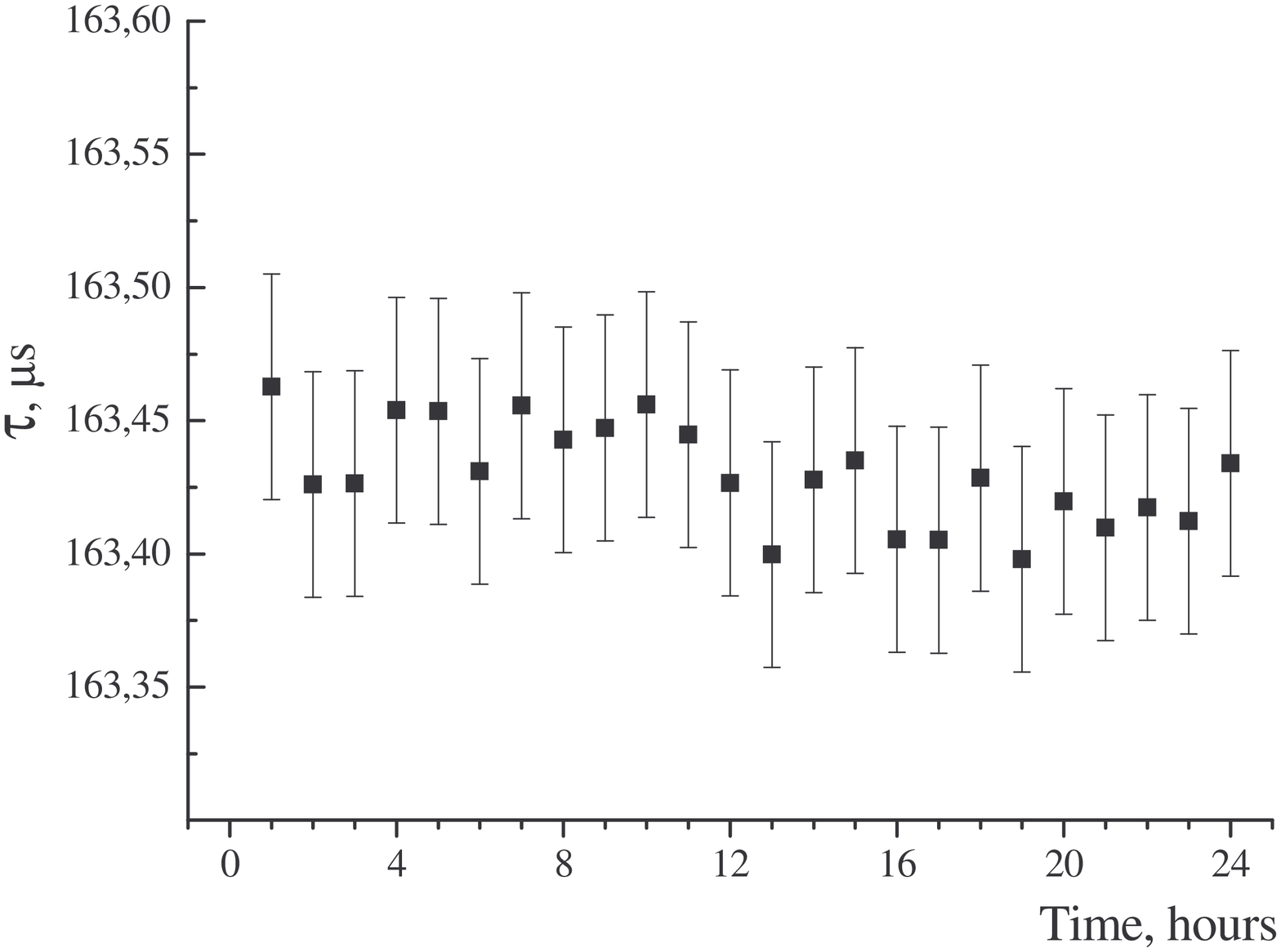}%
\caption{\label{fig13}
Anti-Sidereal-daily variation of half-life of $^{214}$Po obtained by means of moving-average method.
($d_{\rm asd} = d+\Delta t$ = 24 h 3 min 55.9 s).
}
\end{figure}

The possibility of a stochastic realization of sidereal variation was tested also. An analysis was repeated for an artificial anti-sidereal day. Its duration in the solar time is equal to 24 h 3 min 55.9 s.
The result of the analysis is shown on the Fig.\ref{fig13}.
Any variation does not seen within the statistical errors. It gives a confidence that the sidereal daily variation is present really in the data. It is difficult to correlate the sidereal daily variation with any cosmic object at present time because of a nature of the possible influence is unknown. Nevertheless, if such object is exists than it should be an annular variation connected with it. The question needs an additional further investigation.

\section{Results and discussion}

The moving-average method was used to search for annular and daily variations because of a need to improve a statistic of analyzed decay curves and to increase multiply a sensitivity of the analysis to a value of a possible variation. As it seen from a comparison the data on Fig.\ref{fig3} and \ref{fig4}, a value of a statistical error was decreased at $\sim5$ times by increasing of a data accumulation interval from the one week up to 26 weeks (0.5 year). Using of the Fourier-method and wavelet-method to the analysis is complicated by a relatively short size of the time series. Besides, a time $\tau$-series could include not only annular variation but semiannual, for example.
The integrated dependence lost the unambiguity in this case. The used method targeted at a search for the specific variation is determined a univocal binding exactly. It is necessary to compare obtained set of the $\tau$-values with similar data measured independently on the other similar set-up to find a confidence that the variations are nonrandom. Such measurements are carried out in the BNO INR RAS with the TAU-1 set-up since May of 2014. The set-up is placed in the low background laboratory "CAPRIZ" at the 1000 m w.e. depth. The statistics needed to carrying out of a comparison of coincidence behavior of the $\tau$-values time sets is accumulates. It is clear that an essential improvement of statistics could be achieved by a multiple increasing of the dataset rate. The quadratic increase of a random coincidence fraction will occur for the used $^{214}$Po source in a case of a considerable growth of its activity. The value of random coincidence is equal to 1\%
at 12~s$^{-1}$ dataset rate. Such large value connected with a high total activity of all $^{226}$Ra daughter elements in the source and relatively long half-life of $^{214}$Po. Because of it, an increase of dataset rate for the $^{214}$Po without relative growth of a random coincidence background is possible by means of increasing of the number of independent measuring set-ups. This variant seems as a hardly feasible. Another possibility could be realized by using of pair of radioactive isotopes with a similar decay scheme but a shorter half-life. The pair $^{213}$Bi--$^{213}$Po ($T_{1/2}=4.2$~$\mu$s) is a good candidate for such source. This isotopes are the daughter products of $^{229}$Th ($T_{1/2}=7340$~y) \cite{a10} which would be used as a generator isotope. The preparation of the new set-up for $^{229}$Th is carry out at the BNO INR RAS at present time.

\section{Conclusions}

The results of analysis of the data obtained with TAU-2 set-up at the new step of measurements are shown on the presented work. The set-up is  intended to carrying out of a long duration control at a value the $^{214}$Po half-life constant.  The methods of measurement and processing of collected data are reported. Results of the analysis of time series values of $\tau$ with different time step are presented.  Total time of measurements was equal to 730 days. Averaged at 590 days value the $^{214}$Po half-life was found $T_{1/2}=163.46\pm0.04 ~\mu{\rm s}.$
It is shown that the constant feels the daily and annular variations of the unknown nature.
The annual variation with an amplitude $A=(8.9\pm2.3)\cdot10^{-4}$, solar-daily variation with an amplitude $A_{So}=(7.5\pm1.2)\cdot10^{-4}$, lunar-daily variation with an amplitude $A_L=(6.9\pm2.0)\cdot10^{-4}$ and sidereal-daily variation with an amplitude $A_S=(7.2\pm1.2)\cdot10^{-4}$ were found in a series of $\tau$ values. The maximal values of amplitude are observed at the moments when the projections of the installation Earth location velocity vectors toward the source of possible variation achieve its maximal magnitudes. The measurements are continuing.

The work was made in accordance with INR RAS and V.N.Karazin KhNU plans of the Research and Developments.


\begin{thebibliography}{99}
\bibitem{a1}  \textit{Hardy J.C., Goodwin J.R. and Iacob V.E.} //
              {Appl.Radiat.Isot.} {2012. V.70.} P.1931; arXiv:1108.5326 [nucl-ex]; \\
              {doi: 10.1016/j.apradiso.2012.02.021.}

\bibitem{a2} \textit{Bellotti E. et al.} //
            {Phys. Lett.} {B. 2012. V.710.} P.114;
            arXiv:1202.3662 [nucl-ex]; \\
            {doi: 10.1016/j.physletb.2012.02.083.}

\bibitem{a3} \textit{Bellotti E. et al.} //
            {Astropart.Phys.} {2014. V.61.} P.82; 
            arXiv:1311.7043 [astro-ph.SR]; \\ {doi: 10.1016/j.astropartphys.2014.05.006.}

\bibitem{a4} \textit{Jenkins J.H. et al.} //
             {Astropart. Phys.} {2009. V.32.} P.42; 
             arXiv:0808.3283  [astro-ph]; \\ {doi: 10.1016/j.astropartphys.2009.05.004.}


\bibitem{a5}  \textit{Sturrock P.A.,  Fischbach E., and Jenkins J.} //
             {ApJ} {2014. V.794.}  P.42; arXiv:1408.3090 [nucl-th]; \\ {doi: 10.1088/0004-637X/794/1/42.}

\bibitem{a6}  \textit{Sturrock P.A. et al.} //
            {Astropart. Phys.} {2014. V.50.} P.47; \\ {doi: 10.1016/j.astropartphys.2014.04.006.}

\bibitem{a7}  \textit{Nahle O., Kossert K.} //
            {Astropart. Phys.} {2015. V.66.} P.47;  arXiv:1408.5219 [nucl-ex]; \\
            {doi: 10.1016/j.astropartphys.2014.11.005}

\bibitem{a8} \textit{Alexeyev E.N. et al.} //
             {Astropart. Phys.} {2013. V.46.} P.23;  arXiv:1112.4362 [nucl-ex]; \\
             {doi: 10.1016/j.astropartphys.2013.04.005.}

\bibitem{a9} \textit{Wu S.-C.} //
             {Nucl. Data Sheets} {2009. V.110.} P.681; \\
             {doi: 10.1016/j.nds.2009.02.002.}

\bibitem{a10} {Table of Isotopes},
              {Seventh Edition, Edited by Firestone R.B. et al., 8th ed.} {Willey, New York 1996}.

\bibitem{a11} \textit{Gopych P.M. and Zaljubovsky I.I. } //
             {Fiz. Elem. Chast. Atom. Yadra.} {1988. V.19.} P.785.


\bibitem{a12} \textit{Fonda L. et al.} //
              {Rep. Prog. Phys.} {1978. V.41.} P.587; \\
              {doi: 10.1088/0034-4885/41/4/003.}

\bibitem{a13} \textit{Khalfin L.A.} //
              {Physics-Uspekhi} {1990. V.160. No.10.} P.185.

\bibitem{a14} \textit{Misra B. and Sudarshan E.C.G.} //
              {J. Math. Phys.} {1977. V.18.} P.756.

\bibitem{a15} \textit{Facchi P. and Pascasio S.} //
              {J. Phys. A: Math. Theor.} {2008. V.41.} P.493001; \\
              {doi: 10.1088/1751-8113/41/49/493001.}

\bibitem{a16} \textit{Itano W. et al.} //
              {Phys. Rev.} {A. 1990. V.41.} P.2295; \\
              {doi: 10.1103/PhysRevA.41.2295.}

\bibitem{a17} \textit{Alexeyev E.N. et al.} \\
    {Physics of Particles and Nuclei.} {2015. V.46, No.2.} P.157;
    arXiv:1404.5769 [nucl-ex]; \\
    {doi: 10.1134/S1063779615020021.}

\bibitem{a18} \textit{Gavriljuk Ju.M. et al.} //
            {Nucl. Instr. Meth.} {A 2013. V.729.} P.576; arXiv:1204.6424 [physics.ins-det]; \\
            {doi: 10.1016/j.nima.2013.07.090.}

\bibitem{a19} \textit{Medvedev M.N.} //
              {"\emph{Report from gravity works in the Baksan Valley in 2013}".}
              {Report SAI, Moscow, November 2013.}

\bibitem{a20} \textit{Bellini G. et al. (BOREXINO Collaboration)} //
              {Eur. Phys. J.} {A 2013. V.491.} P.92;  arXiv:1212.1332 [nucl-ex]; \\ {doi: 10.1140/epja/i2013-13092-9.}

\bibitem{a21} \textit{Seidelmann P.~Kenneth (ed.)},
            {Explanatory Supplement to the Astronomical Almanac, (at page 698).}
            {United States Naval Observatory.} {Nautical Almanac Office,Great Britain.} 1992.

\end{thebibliography}
\end{document}